\documentclass[prx,aps,twocolumn,showpacs]{revtex4-1}
\draft 
\usepackage{graphicx}
\usepackage{threeparttable}
\usepackage{amsmath}
\usepackage{amssymb}
\usepackage[top=0.6in, bottom=0.65in, left=0.6in, right=0.6in]{geometry}
\begin{document}

\title{The likely determines the unlikely}

\author{Xiaoyong Yan$^{1,2}$, Petter Minnhagen$^{3,}$}
\email{Petter.Minnhagen@physics.umu.se}
\author{Henrik Jeldtoft Jensen$^{4}$}
\affiliation{
$^{1}$Systems Science Institute, Beijing Jiaotong University, Beijing 100044, China\\
$^{2}$Big Data Research Center, University of Electronic Science and Technology of China, Chengdu 611731, China\\
$^{3}$IceLab, Department of Physics, Ume{\aa} University, 901 87 Ume{\aa}, Sweden\\
$^{4}$Centre for Complexity Science and Department of Mathematics, Imperial College London, South Kensington Campus, SW7 2AZ, United Kingdom
}

\begin{abstract}
We point out that the functional form describing the frequency of sizes of events in complex systems (e.g. earthquakes, forest fires, bursts of neuronal activity) can be obtained from maximal likelihood inference, which, remarkably, only involve a few available observed measures such as number of events, total event size and extremes. Most importantly, the method is able to predict with high accuracy the frequency of the rare extreme events. To be able to predict the few, often big impact events, from the frequent small events is  of course of great general importance. For a data set of wind speed we are able to predict the frequency of gales with good precision. We analyse several examples ranging from the shortest length of a recruit to the number of Chinese characters which occur only once in a text. 
\end{abstract}

\maketitle 

\section{Introduction} \label{sec:1}

 A detailed understanding of the mechanisms controlling a certain phenomena can often lead to reliable predictions of what to expect. When one considers complex phenomena, say the weather or language, such a very detailed level of description is typically not possible. Despite of this lack of detail it can still be possible to establish a statistical accurate account of possible behaviours \cite{sornette2000,chialvo2010}. The maximum entropy, or likelihood, principle can be applied to a very broad range of phenomena \cite{jaynes1957,jaynes2003,cover2006,harte2011,bokma13,lee12,baek11a}. The method consists in estimating the probabilistic description, which is statistically most likely to be consistent with the observations available. It is important to keep in mind that no causal mechanistic description is invoked. Rather one assume that  the underlying combinatorial multitude will make happen what is most plausible under given observed constraints. That is, the macro-events generated by the largest number of micro-events are most likely to occur. Say, throw two dice, it is more likely that the sum of the eyes is equal to 7 than equal to 2, since 6 micro-events leads to 7 eyes and only one to 2 eyes.
 
It is therefore to be expected that the method will work for stochastic phenomena like lotteries or dice games. However, in the present work we demonstrate that even for causal highly interdependent and deterministic situations the maximum entropy principle leads often, but not always, to predictions of high precision. Below we will comment on the conditions under which reliable predictions may be expected.

The maximum entropy method combined with Bayesian inference is very well established and used routinely, see ref. \cite{jaynes2003}. Here we describe how the methodology can be developed to obtain accurate estimates of the entire distribution and predictions about extreme behaviour based on just three numbers: a measure of the total ``size", the number of elements and a single measure of most frequent events or the extreme observed. To emphasise the broad applicability we study six different phenomena: heights of humans, wind speed, bus trips, car drives and English and Chinese language. 

In many applications rare extreme events are of particular importance, e.g. gales; while their rarity makes it difficult to estimate their frequency of occurrence from observation of the past. We therefore focus on how the method can extract the statistics of the unlikely from the easily observed most frequent events.

Table \ref{tab:1} show how we in all the considered cases from only three observables, all which are typically easy to access, are able to extract good predictions about various types of extreme or marginal behaviour. One may wonder how a stochastic procedure like the MaxEnt analysis underlying the predictions in Table 1 is able to handle presumably rational and fairly deterministic phenomena like the choice made when traveling a certain distance or the words chosen to express thoughts in a written text. The conclusion is obviously not that some unrecognised stochasticity is in reality governing our choice of words, our travel needs or the growth of recruits or the speed of the wind.  

The reason is that despite each individual choice of journey, expression in terms of words or growth of a person may very well be entirely deterministic, in each case large numbers of possible choices exist which leads to a huge number of combinations. So when considering an large collection of realisations of these choices, we cannot distinguish between underlying proper stochastic processes or deterministic processes with a very large sample space. The situation isn't very different from when we use statistics to analyse the throw of dice.  Each throw is controlled by deterministic mechanics, different throws are subject to slightly different conditions and therefore a large set of throws manifests the combinatorial possibilities available to each deterministic throw. 

Section \ref{sec:2} gives a brief review and motivation of the predictive method by which the results are obtained. Results for six explicit examples are given and discussed in some detail in section  \ref{sec:3}. Finally, a sum-up and a broader perspective is given in section \ref{sec:4}. 

\

\

\

\section{Predictive method} \label{sec:2}

To recall how it is typically applied we consider a set of boxes containing $N$ balls \cite{reif1965}. There are  $N$ boxes and $M$ unnumbered (indistinguishable) balls. The balls are scrambled by randomly picking two boxes and then moving one ball from the first  to the second. The scrambling will produce $N(k)$ boxes which contains $k$ balls. A stationary probability distribution, $P(k)$, describing an ensemble of boxes and balls, is reached after many swaps of balls. In this ensemble the average number of boxes with $k$ balls is given by $NP(k)$. The Shannon entropy of the probability distribution is given by the functional $S[P(k)]=-\sum_kP(k)\ln(P(k))$. To find the most likely distribution $P(k)$ subject to the relevant constraints, one maximises the functional $G[P(k)]=S-b<k>-b<1>$, which imposes the constraint  $M/N=<k>=\sum_kkP(k)$, i.e. the average number of balls in a box together with normalisation condition $<1>=\sum_kP(k)=1$. The result is an exponential $P(k)=A\exp(-bk)$ where $A$ and $b$ are determined from the two constraints.

We can also describe the above ball and box system from an information theoretic view point. Since there are $N(k)$ boxes containing $k$ balls we note that the information needed to determine the specific box a ball ends up in is $\log_2N(k)$ in bits and $\ln(N(k))$ in nats (natural logarithms). Thus the average information needed to associate a ball with a box is $I[P(k)]=\sum_kP(k)\ln(N(k))=\sum_kP(k)\ln(P(k)) + ln(N)$. 
Consequently minimizing $I[P(k)]$ is equivalent to maximizing $S[P(k)]$. Minimizing $I[P(k)]$ subject to appropriate constraints is the key in the present approach. The most general form is \cite{baek11,adamic}
\begin{equation}
G[P(k)]=I[P(k)] + a \langle 1\rangle +b\langle k\rangle +cS[P(k)]
 \label{Eq_G}
\end{equation}
where 
\begin{equation}
\begin{aligned}
I[P(k)]=\sum_kP(k)\ln(g(k)N(k)\\
=\sum_kP(k)\ln(g(k)P(k)) + ln(N)
\end{aligned}
 \label{Eq_I}
\end{equation}

Combinatorial information concerning the order in each box is contained in the function $g(k)$. In the example above one has $g(k)=1$ (because the balls are unnumbered and only the boxes are labeled). $G[P(k)]$ is minimized subject to the two constraints normalization and average. A second example elucidates this: Suppose the balls are numbered and randomly put one by one into the boxes. If a box ends up with three specific balls numbered 1, 2, and 3, then there are 3!=6 different orders in which they could have arrived into the box. This means that the system by definition contains the information to distinguish these 6 possibilities. Thus $I[P(k)]$ changes to $I[P(k]=\sum_kP(k)\ln(k!)P(k)) + ln(N)$ so that  $g(k)=k!$ in this case. Minimizing $G$ in Eq.(\ref{Eq_G}) constraint now instead gives a Poisson distribution $P(k)=A\exp(-bk)/k!$ \cite{reif1965}. The point is that both of the examples are characterized by an information function $g(k)$ associated with a box of size $k$. Suppose that $g(k)$ is \textit{unknown} but one instead knows the entropy $S$. Then one can obtain a condition for $g(k)$ by minimising $I$ subject to the additional constraint of fixed entropy $S$. This is the content of Eq.(\ref{Eq_G}). Thus the strategy is to find an approximation for the unknown combinatorial information characteristics of a system from a known entropy. This is the opposite from the more conventional use of maximum entropy, in which characteristic probability density functions are determined from the condition that they maximise the entropy.

In order to achieve this one determines, for a given system, what the basic description of the system corresponds to in terms of a random grouping process. All additional information about the system will be incorporated as constraints within the maximum entropy approach. In the examples of Table \ref{tab:1} one knows that the basic entities,  i.e. recruits, wind-speed readings, bus-trips, car-drives, words and Chinese characters, are sorted into to groups i.e. length-groups, wind-speed groups, mileage-groups, groups of words with the same spelling, groups of Chinese characters identically drawn. We now do the combinatorics or information analysis of these groups. For a  group of size $k$ the corresponding group information is $\ln(g(k))=\ln(k)$ i.e. the information needed to identify an entity within the group. If this is all that is known then the maximum entropy  corresponds to the situation when all the entities are equally likely to be assigned any of the total possible $M=\sum_kkNP(k)$ group-labels. If one in addition knows the average $\langle k\rangle$, then minimisation of Eqs.(\ref{Eq_G}) and (\ref{Eq_I}) yield the form of the most likely distribution $P(k)=A\exp(-bk)/k$. Furthermore, if in addition the entropy is known the most likely distribution acquires the functional form 
\begin{equation}
P(k)=A\exp(-bk)k^{-\gamma}
\label{Eq_RGF}
\end{equation}
where the three constants $A$, $b$ and $\gamma$ are determined from the known data. We use below Eq. (\ref{Eq_RGF}) as an ansatz when analysing data sets.  

The explicit steps are: Minimising the the functional $G[P(k)]$ with respect to $P(k)$ where 
\begin{equation}
G[P(k]=\sum_kP(k)\ln(kP(k))+ a \langle 1\rangle +b'\langle k\rangle +cS[P(k)]
 \label{Eq_GE}
\end{equation}
by solving the minimum condition given in terms of the functional derivative $dG[P(k)]/dP(k)=0$, which leads to  
\begin{equation}
\ln(kP(k)) +1 +a +b'k -c\ln(P(k))-c=0
\end{equation}
or 
\begin{equation}
\ln(k^\gamma A^{-1} \exp(bk)P(k))=0
\end{equation}
where $\gamma=1/(1-c)$, $b=b'/(1-c)$ and $A=\exp(-1-a/(1-c))$ from which Eq.(\ref{Eq_RGF}) follows.

\begin{figure*}[!t]
\centering
\includegraphics[width=1.0\textwidth]{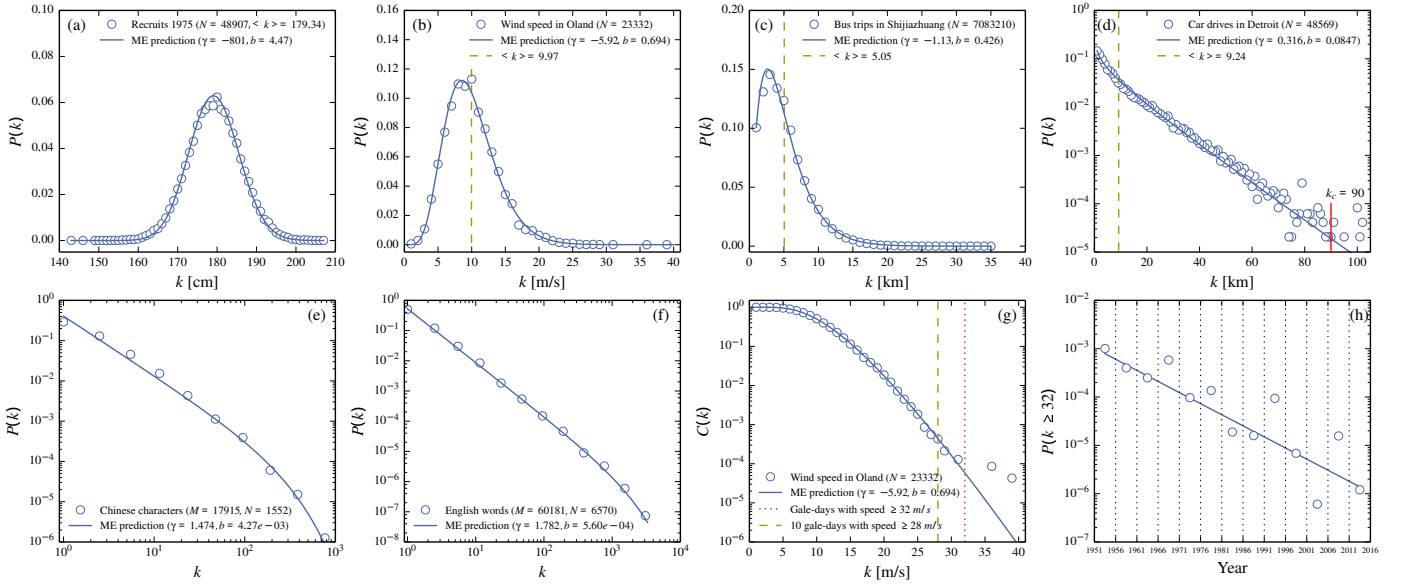}
	\caption{\textbf{Likelihood predictions.} Data are given by open circles and the predictions by the full curves. The full curves in (a)-(f) are all given by the predicted form $P(k)=A\exp(-bk)k^{-\gamma}$ where the parameters $A$, $\gamma$ and $b$ are determined by minimal information, as described in the text. The subfigures (a)-(f) corresponds to the six cases: height of recruits; wind speed (maximum wind for each day taken as the average over 10 minutes at regular intervals. The weather station is placed at the south tip of \"{O}land); bus-travels in the city Shijiazhuang; car drives in Detroit; frequency of Chinese signs in a text; word-frequency in an English text, respectively. The agreement is dramatic. Subfigure (g) shows the wind-data presented in cumulant form giving the probability for an event larger than the value on the x-axis. The dashed vertical line corresponds to the prediction of 10.3 gale-days $\geq 28$ m/s during the measuring period 1951-2015. The observed number is 10. The dotted vertical line is for gale-days $\geq 32$ m/s where the prediction is 1.38 days and the observed 2. Finally, subfigure (h) shows a linear regression (straight line) for the predictions of each five year period (open circles). The regression line suggest that the risk for gales of this size have declined of the order of a factor 100 over the observation period.
} 
	\label{fig1}
\end{figure*}

To sum up: we are essentially maximising the likelihood, or entropy, under the appropriate constraints given by in the $M$, $N$ and $k*$ columns in Table \ref{tab:1}. However, since the missing information in the examples are basically the underlying emerging stochasticities, we invert the procedure by instead determining the effective stochasticity which will have a given entropy as its maximum. Under broad conditions this leads to the unique probability distribution $P(k)=A\exp(-bk)k^{-\gamma}$ which simultaneously describe all the different distributions in Fig. \ref{fig1}.

\begin{table*}[!t]
\caption{Likelihood Predictions.}\label{tab:1}
\begin{tabular*}{18.5cm}{@{\extracolsep{\fill}}|c|c|c|c|c|c|c|}
\hline
Known Data  & $M$ & $N$ & $k*$ & Prediction & Obtained & Measured \\
\hline
Recruits$^1$ & 8770975 cm & 48907& $k_{max}=207$ cm & $\Longrightarrow$ & 74.88\% & 69.08\%\\
\hline
Wind speed$^2$ &232695 m/s & 23332&  5.5\% is 5 m/s& $\Longrightarrow$ & 0.0058\% & 0.0084\%\\
\hline
Bus-trips$^3$ &35770210 km &7083210 & 10\% $\in[0,1]$ km & $\Longrightarrow$ & $\in[2,3]$ km& $\in[2,3]$ km\\ 
              &             &          &                 &$\Longrightarrow$ &$\geq 9$ km=10\%  & $\geq 9$ km=9.8\%\\
\hline
Car-drives$^4$ &448778 km & 48,569& $k_c=90$ km & $\Longrightarrow$ & 14.7\%& 14.5\%\\
\hline
Signs$^5$ &17915 &1552 & $k_{max}=747$& $\Longrightarrow$ &  40.51\% &  29.12\% \\ 
\hline
English$^6$ & 60181&6570& $k_{max}=3300$& $\Longrightarrow$ & 50.44\% & 52.69\% \\ 
\hline
\end{tabular*}
\begin{tablenotes}
        \footnotesize

\item[1] Swedish recruits born 1975 (see Section \ref{sec:3}): $M$ = total length, $N$ = total number, $k_{max}$ = tallest, $\Longrightarrow$ obtained = shortest in \% of tallest. 
\item[2] Wind speed in \"{O}land island, Sweden (see Section \ref{sec:3}): $M$ = total wind speed observed, $N$ = total observation days, $k*=5.5$\% of days the observed wind speed is 5 m/s, $\Longrightarrow$ obtained = \% of wind speed equal to or larger than 32 m/s. 
\item[3] Bus-trips in Shijiazhuang (see Section \ref{sec:3}):  
$M$ = total distance, $N$ = total number of trips, $k*=10$\% of trips in interval [0,1] km, $\Longrightarrow$ obtained = position of maximum and \% of trips larger than 9 km. 
\item[4] Car-drives in Detroit (see Section \ref{sec:3}): 
$M$ = total distance, $N$ = total number, $k*=$ the longest 10 trips longer than $k_c=90$ km, $\Longrightarrow$ obtained = \% of trips in interval [0,1] km. 
\item[5] The Chinese novel {\it A Q Zheng Zhuan }by Xun Lu written in Chinese characters (see Section \ref{sec:3}):  
$M$ = total number of characters, $N$ = number of different characters, $k_{max}=$ occurence of most the frequent character, $\Longrightarrow$ obtained = \% of characters occurring only once. 
\item[6] The English novel {\it Under the Greenwood Tree} by Thomas Hardy (see Section \ref{sec:3}):
$M$ = total number of English words, $N$ = number of different words, $k_{max}=$ occurence of most the frequent word, $\Longrightarrow$
obtained = \% of words occurring only once. 

\end{tablenotes}

\end{table*}

\begin{table*}[!t]
\caption{Pairwise Likelihood Predictions.}\label{tab:2}

\begin{tabular*}{18.5cm}{@{\extracolsep{\fill}}|c|c|c|c|c|c|c|}
\hline
Known Data  & $M$ & $N$ & $k*$ & Prediction & Obtained & Measured\\
\hline
Recruits 1$^1$ & 8770975& 48907& $k_{max}=207$ cm& $\Longrightarrow$ & 74.88\%& 69.08\%\\
\hline
Recruits 2$^2$ &10516502 cm& 58698 & $k_{max}=211$cm& $\Longrightarrow$ & 72.04\%& 65.40\%\\
\hline 
Wind 1$^3$  & 232695 &23332  & 5.5\% is 5 m/s & $\Longrightarrow$ & 0.0058\% &0.0084\%  \\
\hline 
Wind 2$^4$   &236968  &23408  & 5.0\% is 5 m/s & $\Longrightarrow$ &0.0150\%  & 0.0085\% \\
\hline
Car-drives Detroit$^5$ &448778 km& 46541 & $k_c=90$ km& $\Longrightarrow$ & 14.7\%& 14.5\%\\
\hline
Car-drives Seattle$^6$ &525571 km&65861 & $k_c=83$km& $\Longrightarrow$ & 16.89\%& 17.94\%\\ 
\hline
\end{tabular*}
\begin{tablenotes}
        \footnotesize
\item[1] Recruits 1 is the Swedish recruits born 1975 shown in Fig. \ref{fig1}(a).
\item[2] Recruits 2 is the year-group born 1965 shown in Fig. \ref{fig2}.
\item[3] Wind 1 is the observations from \"{O}lands south tip during the period 1951-2015 shown in Fig \ref{fig1}(b).
\item[4] Wind 2 is the observations from Svenska H\"{o}garna during the period 1951-2015 shown in Fig. \ref{fig4}(a).
\item[5] Car-drives in Detroit shown in Fig. \ref{fig1}(c).
\item[6] Car-drives in Seattle shown in Fig. \ref{fig5}.
 \end{tablenotes}
  \end{table*}

\begin{figure}[!t]
\centering
\includegraphics[width=0.45\textwidth]{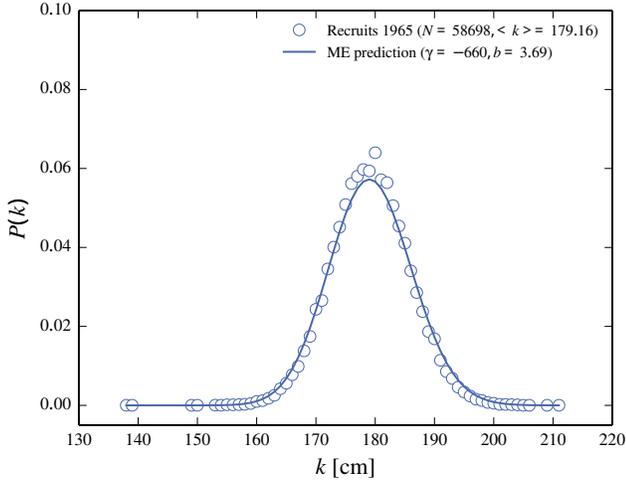}
\caption{\textbf{Recruit 2.}
Height of recruits: Full curve= MaxEnt prediction based on the three numbers $( M,N,k_{max})$; data=open circles.} 
\label{fig2}
\end{figure}

\begin{figure*}[!t]
\centering
\includegraphics[width=0.7\textwidth]{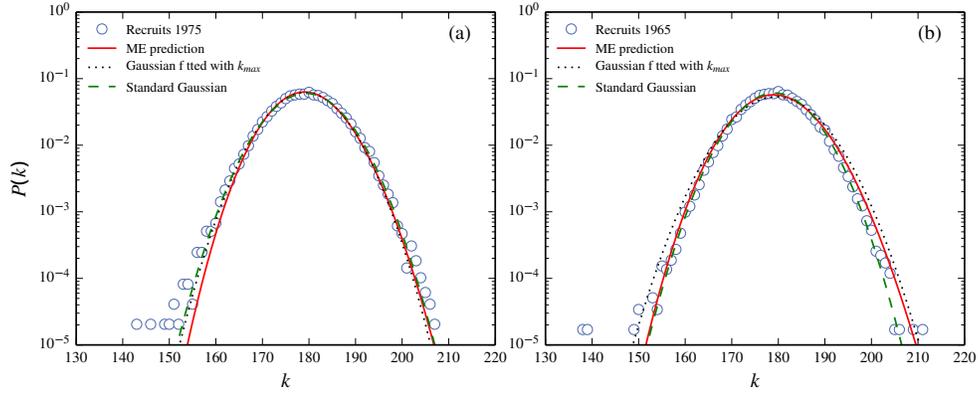}
	\caption{\textbf{Recruit 1 and Recruit 2 data and predictions in lin-log plot.}
	Recruit 1 are recruits born 1975 and recruit 2 are born 1965. The full curves are the MaxEnt-predictions based on the three numbers
	$( M,N,k_{max})$ given in the table. The dashed curves are standard fits to Gaussian distributions and the dotted curves are Gaussian distributions which have the same $k_{max}$ as the data. The predicted ratios Shortest/Tallest are for Recruit 1: 69\%, 75\%, 73\% and 73\%, for data, MaxEnt-, standard Gaussian, and $k_{max}$-fitted Gaussian, respectively. For Recruit 2: 65\%, 72\%, 74\%, 71\% .}
\label{fig3}
\end{figure*}

\begin{figure*}[!t]
\centering
\includegraphics[width=1.0\textwidth]{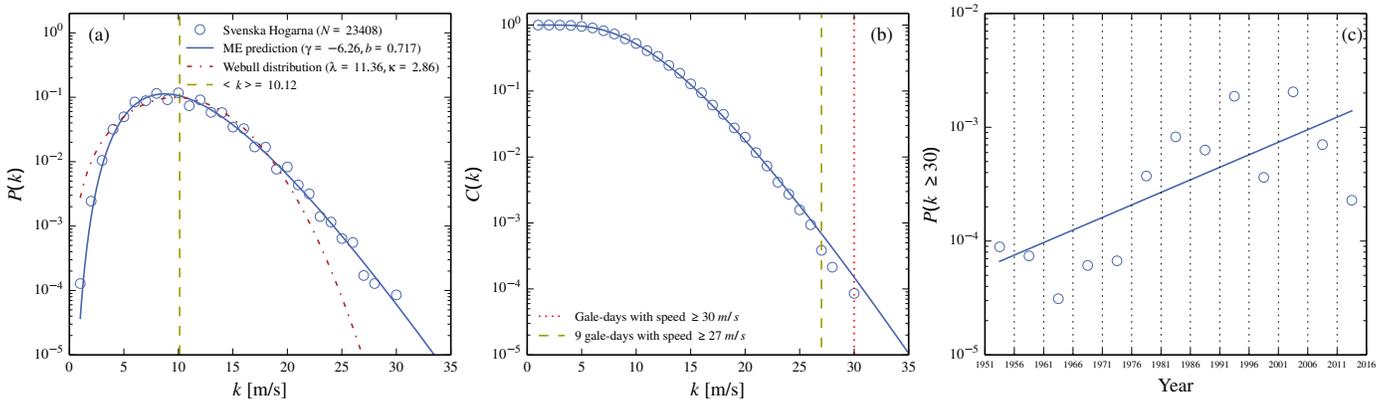}
	\caption{\textbf{Wind-data from Svenska H\"{o}garna.}
	Data from a weather station about 500 km to the north of \"{O}land south tip on the Swedish coast for the period 1951-2015. The figures correspond to Fig. \ref{fig1} (b), (f), and (g). Subfigure (a) demonstrates the striking agreement between observations and the MaxEnt prediction based on the minimal info given in Table \ref{tab:2}. The dashed-dotted line is the two parameter Weibull distribution ($W(k)=\frac{\kappa}{\lambda}(\frac{k)}{\lambda})^{\kappa-1}e^{-(k/\lambda)^\kappa}$) commonly used in analyzing wind-data \cite{morgan2011}. For the same info (probability for 5 m/s and average speed) the MaxEnt predicts strong winds with much higher accuracy. 
Subfigure (b) illustrates the same data and MaxEnt-prediction in cumulant form. The dashed horizontal line gives the prediction for $\geq 27$ m/s gale-days and the dotted for $\geq 30$ m/s gale-days. The predictions are 16.1 respectively 3.5 days (of a total of about 23000 days) whereas the corresponding observations are 9 and 2, respectively. Thus the predictions for these extreme rare events are within a factor of two of the actual observations. Subfigure (c) approximate the MaxEnt predictions for consecutive five year periods from 1951-2015 with a linear regression. The straight line suggests that the risk for extreme winds have increased from 1970 to 2005.}	
\label{fig4}
\end{figure*}

\begin{figure}[!t]
\centering
		\includegraphics[width=0.45\textwidth]{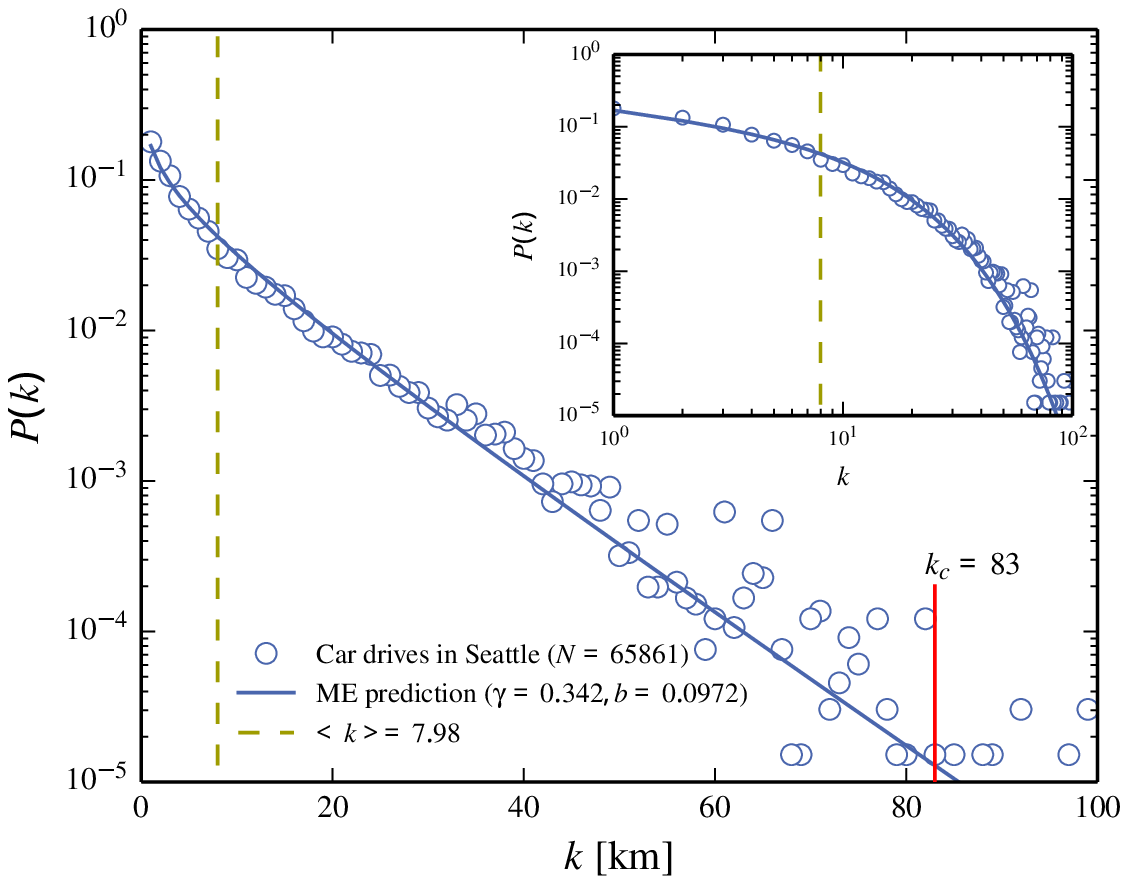}
	\caption{\textbf{Car drives in Seattle.}
	Data=open circles; Full curve=MaxEnt-prediction based on the three numbers $(M,N,k_c)$ given in Table \ref{tab:2}. $k_c$ is the distance which is exceeded by only ten of the about 50000 drives (vertical full line). The average distance is denoted by the vertical dashed line. Note the up-bend in the data and prediction towards smaller distances in the lin-log-plot. The prediction is that about 17\% of the trips are in the smallest interval [0,1] km. The actual value is 18 \%.  The insert shows the same data and prediction in a log-log-plot} 
	\label{fig5}
\end{figure}

\section{Predictions for Six Different Real Systems} \label{sec:3}

Figure \ref{fig1} (a)-(f) display six very different systems which all are well predicted by the MaxEnt-distribution ansatz $P(k)=A\exp(-bk)k^{-\gamma}$. The difference in shape when going from a) to f) is essentially related to a steady increase of $\gamma$ from -800 to 1.8.

\textbf{\textit{Length distribution}:}
The first example (Fig. \ref{fig1}(a)) concerns the length distribution of Swedish recruits. The data consists of the height of Swedish men drafted into the Swedish military at the age of 18. Recruit 1 is men born 1965 and Recruit 2 1975. The data is obtained from the Swedish drafting authority. From the knowledge of the number of recruits (48907), their average length (179.3 cm) and the length of the tallest recruit (207cm) we estimate the likely length of the shortest. The prediction based on maximum entropy is given in Table \ref{tab:1}: the shortest recruit is predicted to be 75\% of the height of the tallest and the actual value is 69\%. Note that the distribution of the heights is a priori unknown, but an estimated is obtained through the predictive method from the first three numbers given in Table \ref{tab:1}. Despite of only using three observed numbers and the ansatz in Eq. (\ref{Eq_RGF}) the prediction is of good accuracy: out of the about 50000 recruits only about 0.03\% are smaller than the predicted value. Suppose you {\it in addition} used the not unreasonable assumption that the heights are following a normal distribution \cite{quetelet1942}. This distribution is symmetric around the average length, so you may predict that the shortest person is around 73\% . The additional knowledge apparently improved the prediction very little. A second example of recruits are given in Fig. \ref{fig2}. The three numbers, on which the MaxEnt-prediction is based, are given in Table \ref{tab:2}. Figure \ref{fig2} shows that the MaxEnt-prediction works equally well for the data-set Recruit 2 as for Recruit 1 (compare Fig. \ref{fig1}(a)). In Figure  \ref{fig3} the two data sets for the recruits are shown in lin-log plots together with a standard Gaussian-fits to the data, as well as the $k_{max}$-fitted Gaussians. One may note that ratios shortest/tallest compared to the actual ratios are best predicted by the $k_{max}$-fitted Gaussian. However, the differences are small (the actual values are given in the caption of Fig.  \ref{fig3}).

\textbf{\textit{Wind Speed Distribution}:}
The data is obtained from the weather-stations at the south tip of \"{O}land (1951-2015) and Svenska H\"{o}garna (1951-2015). The wind is measured regularly (typically each third hour. Each observation is the wind-speed average over ten minutes. The data analyzed here is the maximum recording for each day, from SMHI (the Swedish meteorological and hydrological institute,  http://opendata-catalog.smhi.se). This set of data is analysed in Fig. \ref{fig1}(b)). As input we use the total number of days (23332), the average wind speed (9.97 m/s) and the frequency of 5 m/s winds (5.5 \% of the days). From these three numbers one predicts that 1.4 days with windspeed recording $\geq$ 32 m/s are expected for this period. The actual number is 2 (compare Table \ref{tab:1}). This is an example of how the likely determines the unlikely: the recording of a wind-speed of 5 m/s is a likely event, for which good statistics can be obtained. From the knowledge of these likely events and the knowledge of the average wind-speed, our method is able to predict the extremely unlikely recordings of $\geq 32$ m/s winds to better than a factor of two. Fig. \ref{fig4} and Table \ref{tab:2} row 4 gives a second example from a weather station at Svenska H\"{o}garna for the same period. The MaxEnt prediction is again quite accurate (see Fig. \ref{fig4}(a)). The corresponding cumulant prediction is given Fig. \ref{fig4}(b). The conclusion is the same as for the \"{O}land data: the prediction for the number of gale-days for $\geq 30$ m/s is 3.5 on the average whereas the observed number is 2, which is a deviation of only 30\%. It is interesting to compare the risk change for extreme gales during the period 1951-2015 for the two weather stations given by Fig.  \ref{fig1}(h) and Fig. \ref{fig4}(c): whereas the \"{O}land data suggests a steady decrease by a factor of about 100, the data from Svenska H\" {o}garna suggests an increase within the period 1970-2005 with about a factor of about 10. This might indicate natural cyclic long time changes at the locations or maybe some more permanent climate change.

\textbf{\textit{Length-distribution of Bus-trips}:}
The third example concern the length of bus-travels in the city Shijiazhuang (see Fig. \ref{fig1}(c)) as recorded by transit smart cards. The data is obtained from the Shijiazhuang public transportation corporation \cite{yan2013}. For a data-set consisting of a large number of travels (see Table \ref{tab:1}) the average distance is 5.1 km and the percentage of travels up to 1 km is 10 \%. Based on these two numbers we estimate the most likely distance for a bus-trip as well as the distance traveled by the 10\% of the people, who make the longest rides. One could argue intuitively that it is probably less convenient to take very short rides because you can walk or bike, the buses run infrequently, the fares for shorter rides are too high, or the bus-stops are too infrequent {\it etc}. Accordingly one might expect the most likely traveled distance to be, say, of about the average distance 5.1 km. Perhaps one would a first expect a somewhat symmetrical distribution centred around the average distance in a way to make the 10 \% longest rides longer than about 9 km. However, the MaxEnt predicts something different: The most likely ride is between 2-3 km and the 10\% of the people who make the longest travels go more than 9 km, implying instead a very skewed distribution. The actual values are indeed that the most likely travel is between 2-3 km and that the 10\% of the people, who make the longest travels, go more than about 9 km. The full MaxEnt-prediction is compared to the data in Fig. 1\ref{fig1}(c). Perhaps one can intuitively understand the structure of the distribution based rational human behaviour, the likely organisation of the buses, and the pricing. The fact that many people prefer to walk the first few kilometers rather than finding a bus-stop and wait for a bus, seem intuitively very natural.  Nonetheless the MaxEnt avoids having to make postulates about rational behaviour and manage to make accurate statistical predictions. Does this imply that people do not make rational decisions based on sound reasoning? No, it is the other way around. It is not the lack of reasons which makes maximum entropy work, it is the abundance of possibilities.

\textbf{\textit{Length-distribution of Car-drives}:}
The fourth example (Fig. \ref{fig1}(d)) concerns the length of car-drives in US cities.  Travel survey data of Detroit (1994) and Seattle (2006) obtained from the Metropolitan Travel Survey Archive website (http://www.surveyarchive.org/). The data contains one-day travel diary of sampled households, including each trip's origin and destination, start and end time, trip mode and purpose, from which we extracted the distance of all the car trips. The data set from Detroit includes about 48000 drives with the average length 9.2 km. The longest 10 drives are longer than 90 km. We now ask what is the proportion of the shortest drives up to 1 km? Is it like the bus-trips in Shijiazhuang that it is more convenient to walk or bike the shortest distances? Or does the American habit of driving everywhere combined with the fact that sidewalks are rare in US cities imply that the shortest distances are the most common? The MaxEnt predicts that the shortest distance is indeed the most common and that 14.7\% of the drives are in the interval between [0-1] km. The actual percentage is 14.5\%. A second example, Seattle, is given in Fig. \ref{fig5} with the data in Table \ref{tab:2}. The car-trips in Seattle shows the same feature as the car-trips in Detroit. The MaxEnt-predictions are of good accuracy for both data-sets. One may note that the proportion of drives in the shortest interval [0,1] km is slightly larger in Seattle. The actual percentage is 14.5\% and 17.94\% for Detroit respectively Seattle, whereas the MaxEnt predictions are 14.7\% and 16.89\%, respectively. The agreement with data again leads to the surprise that although we can intuitively understand the structure from the alleged car-driving habits in US, yet the MaxEnt-estimate is able to make accurate predictions without need of such insights. From a causal point of view the connection between shortest and longest trips is somewhat baffling: the drivers who made the 10 longest drives in Detroit (more than 90 km) presumably had good reasons to make these drives, but why should the incentives for these ten longest drives also determine that about 15\% of the drives where shorter than 1 km? 

\textbf{\textit{Distribution of Chinese Characters}:}
The fifth example (Fig. \ref{fig1}(e)) concerns the distribution of Chinese characters in a text: the Chinese novel, {\it A Q Zheng Zhuan} by Xun Lu. We first removed punctuation marks and numbers from the novel, then counted the Chinese characters one by one and finally got the characters frequency distribution. The short story  contains about 18000 Chinese characters of which about 1500 are distinct  and the most common one appears about 750 times (see Table  \ref{tab:1} for precise numbers). Without making use of any information concerning the Chinese language or its representation by Chinese characters, we now estimate how many characters are likely to appear only one time in the novel. The prediction from the three numbers in Table \ref{tab:1} gives about 40\% compared to the actual about 30\% of the total number of different characters in the novel. So the method applied is again accurate to within 25\% despite being purely based on combinatorics and not invoking any linguistic knowledge of structure, grammar and representation of a language \cite{yan2015}.

 \textbf{\textit{Distribution of English Words}:}
The sixth example (Fig. \ref{fig1}(f)) concerns the distribution of words written by letters. We analyse the novel {\it Under the Greenwood Tree } by Thomas Hardy, which  contains about 60000 English words written by letters, about 6500 are distinct and the most common word appears about 3300 times. The proportion of words which appear only once in the novel is predicted to be 50.4\%, whereas the actual proportion is 53.7\%. We see again that information about the most frequent (in this case the word``The", which growths in direct proportion to the length of the text) enables a accurate prediction of the statistics of the infrequent words. 

\section{Conclusions} \label{sec:4}

The high degree of accuracy obtained by the predictive method employed (see Tables \ref{tab:1} and \ref{tab:2} and Fig. \ref{fig1}) demonstrates that the collection of all possible outcomes is statistically well characterized by just a few constraints extracted from the observed collection of data. 

Typically the predictions are within a few percents of the measured values. But there are situations where the analysis is less precise, e.g. the prediction of gales and the analysis of Chinese novel written using Chinese sign language. The latter is related to the high level of degeneracy (multiple meanings assigned to the same Chinese character) of sign language, see ref. \cite{yan2015}. What this means is that the combinatorial multitude characterising words in the Chinese language is not directly represented by the individual characters. Said in another way: the characters acquire their word meaning from the context. Of course this is always to some extend the case but lesser so when words are spelled out using an alphabet. Why is the degeneracy a problem for MaxEnt to work? Simply because the MaxEnt analyses the combinatorial structure of the multitude of possible events. But if the same label is used for different events problems arise. Just think of dice. If both one eye and two eye faces are labeled with one dot, then the counting of outcomes will go wrong.

In the case of wind speed, it might very well be that the dynamics of the atmosphere isn't uniquely described by the daily wind speed. Nevertheless, the MaxEnt-predictions given in Table \ref{tab:2} rows 3 and 4, and in Figs. \ref{fig1} (b), (g) and Fig. \ref{fig3},  are remarkable in view of the fact that the wind data exhibit significant non-stationarity: the estimated average number of strong gale-days ($\geq 32$ m/s) within the period 1951-2015 is 1.4, based on the observations of the frequency for 5-m/s-winds and the average wind-speed. The observed number is 2. The fact that the easily observed can be used to predict rare events makes it possible to estimate how the risk for extreme winds changes over time. This follows because the input parameters, like the frequency of 5-m/s winds and the average wind are rather well determined even over shorter periods. The linear regression in Fig. \ref{fig1}(h) suggests that the risk has decreased by a factor of about 100, from about one in three years at the beginning of 1950, making it extremely improbable to encounter a $\geq 32$m/s-gale at this location during the next five year period. A long trend like this might well be linked to long-time weather cycles or climate changes. 

Fig. \ref{fig4}(a) also compares the present prediction with the two parameter Weibull distribution which is more commonly used to describe wind speed data \cite{morgan2011}. When the same information is used, the MaxEnt-prediction predicts the strong wind distribution much better (see Fig. \ref{fig4}(a)). Thus in this case our method predicts the unlikely from the likely much better and this could well be true for a wider range of risk-estimates for rare events.  

Despite individual travels presumably most often are dictated by rational decisions, the collective statistics of bus-rides or car-rides shown in Table \ref{tab:1} rows 3 and 4 and Figs. \ref{fig1}(c) and (d)  is very well captured by the corresponding MaxEnt distribution. This more generally implies that deterministic system may well display characteristic features that are effectively random.

But could the agreement between the data and our predictive method not just be accidental rather than a proof that randomness serves as an excellent effective statistical description? One case where this can be explicitly answered is the word-frequency distribution in Fig. \ref{fig1}(f): it was shown in ref. \cite{yan2015b}  that the shape of the word-frequency distribution changes shape as a function of the text-length and that furthermore also this change is very well described by use of the MaxEnt procedure discussed in detail in the present paper.

Our conclusions are that some (and possibly many) complex deterministic systems contain what may be termed \emph{pseudo-random} features originating in the huge combinatorial multitude of micro configurations underlying the observed systemic behaviour. Furthermore these pseudo-random features can sometimes be characterised by only a minimal knowledge of the system and the distribution of possible outcomes can be predicted from a MaxEnt-method. The obtained distributions of possibilities can then be used to predict the probability of rare events. 

On the other hand, if very little specific information about the system is needed to derive the distribution, then very little specific information about the system can be deduced from the distribution: You cannot extract any system specific features from the shape of the distribution if it is pseudo-random. In such cases, to understand the system in depth, one will have to resort to correlations or multi-variable probability distributions. E.g. in the case of written text: the frequency distribution may well be considered a pseudo-random object but the frequency of pairs, triplets etc. of words will depend on gramma and meaning.


\end{document}